\documentclass[%
 aip,
 amsmath,amssymb,
 reprint,%
]{revtex4-1}

\usepackage{graphicx}% Include figure files
\usepackage{dcolumn}% Align table columns on decimal point
\usepackage{bm}% bold math
\usepackage[utf8]{inputenc}
\usepackage[T1]{fontenc}
\usepackage{mathptmx}
\usepackage{etoolbox}

\makeatletter
\def\@email#1#2{%
 \endgroup
 \patchcmd{\titleblock@produce}
  {\frontmatter@RRAPformat}
  {\frontmatter@RRAPformat{\produce@RRAP{*#1\href{mailto:#2}{#2}}}\frontmatter@RRAPformat}
  {}{}
}%
\makeatother
\begin{document}

\preprint{AIP/123-QED}

\title[Control of a single-photon router via an extra cavity]{Control of a single-photon router via an extra cavity}
% Force line breaks with \\
\author{Yi-Ke \surname{Luo}}
\affiliation{Synergetic Innovation Center for Quantum Effects and Applications, Key Laboratory for Matter Microstructure and Function of Hunan Province, Key Laboratory of Low-Dimensional Quantum Structures and Quantum Control of Ministry of Education, School of Physics and Electronics, Institute of Interdisciplinary Studies, Hunan Normal University, Changsha 410081, China}
\author{Ya \surname{Yang}}
\affiliation{Synergetic Innovation Center for Quantum Effects and Applications, Key Laboratory for Matter Microstructure and Function of Hunan Province, Key Laboratory of Low-Dimensional Quantum Structures and Quantum Control of Ministry of Education, School of Physics and Electronics, Institute of Interdisciplinary Studies, Hunan Normal University, Changsha 410081, China}
\author{Jing \surname{Lu}}
\affiliation{Synergetic Innovation Center for Quantum Effects and Applications, Key Laboratory for Matter Microstructure and Function of Hunan Province, Key Laboratory of Low-Dimensional Quantum Structures and Quantum Control of Ministry of Education, School of Physics and Electronics, Institute of Interdisciplinary Studies, Hunan Normal University, Changsha 410081, China}
\author{Lan \surname{Zhou}*}
\affiliation{Synergetic Innovation Center for Quantum Effects and Applications, Key Laboratory for Matter Microstructure and Function of Hunan Province, Key Laboratory of Low-Dimensional Quantum Structures and Quantum Control of Ministry of Education, School of Physics and Electronics, Institute of Interdisciplinary Studies, Hunan Normal University, Changsha 410081, China}
\email{zhoulan@hunnu.edu.cn}%Lines break automatically or can be forced with \\

\date{\today}% It is always \today, today,

\begin{abstract}
Controllable single-photon routing plays an important role in quantum networks. We investigate single-photon scattering in two one-dimensional (1D) waveguides by a three-level emitter with a cascade configuration, which is a dipole coupled to an extra cavity. The tunneling path for the transmission of a single photon is switched by whether the extra cavity contains photons. For the setup, the Autler-Townes splitting is modulated by the extra cavity, in which the transmission valley (reflection range) width is tunable in terms of the number of photons in the extra cavity. Our investigation will be beneficial to single-photon routing in quantum networks using quantifiable photon numbers in an extra cavity.
\end{abstract}

\maketitle

Quantum network, which is the essence of quantum information processing, and
the key lies in the ability of quantum state transfer, that is, evaluating
the effectiveness of transferring quantum states from one end to the other %
\cite{1}. It is well known that photons are one of the ideal candidates as information carriers for long-distance optical quantum communication due to their low decoherence rate \cite{2,3,4}. Therefore, how to efficiently and perfectly route
photon signals in quantum channels is an important topic in quantum
communication. At present, the quantum router has been theoretically and
experimentally studied in a variety of systems, i.e., coupled-resonator
waveguides (CRWs) \cite{5,6,7,8,9,10,11},
superconducting circuits \cite{12,13,14,15,16,17}, optomechanical systems \cite%
{18,19,20,21,22,23}, and atomic systems \cite%
{7,24,25,26,27,28,29,30,31,32,33,34}. In addition, using
waveguide-QED systems \cite{35,36}, researchers have put forward many proposals for single-photon (SP) routing \cite%
{5,7,8,10,28,29,37,38,39,40,41}%
. The SP router with multiple channels can be achieved by controlling the driving field strength of the $\Lambda -$type three-level system (3LS) and its coupling strengths with the waveguides \cite{41,42}. It is interesting to study the controllable single-photon transport through the extra cavity. The Autler-Townes splitting is modulated by the extra cavity \cite{15,17,20,21,43,44,45}, and the width of the transmission valley can be adjusted by controlling the number of photons in the extra cavity.

Here, we introduce a scheme for a four-port system, which is based on quantum
interference tuned by chiral waveguide-emitter coupling \cite{42}. In this paper, the
photon transport can be controlled by an extra cavity. If the extra cavity is in a vacuum state, our scheme can be regarded as two independent
one-dimensional (1D) waveguides chiral coupled to a two-level system \cite%
{46}. However, once the extra cavity contains photons, the
input single photon will appear to have more abundant transmission properties. It is shown
that the full transmission of the input single photon can be obtained in the
resonance case. This is the condition of the total reflection of the
two-level system \cite{46}. Moreover, we analyzed the scattering spectrum
of a single photon using the parameters of the extra cavity \cite{47,48,49,50}, such as the
coupling strength to the emitter, photon number, and resonance frequency.
Incidentally, our scheme can be regarded as two 1D waveguide chiral
coupled to a V-type atom in the dressed state representation. In this way, the dressed state makes it easier to understand the physical images than the
bare state. First, we present the four-port device and its Hamiltonian. Introducing the parity operator to divide it into the scatter-free subspaces and the controllable subspaces. Then we study the scattering process for a single photon and explain how to adjust the parameters of the extra cavity to realize perfect single-photon routing.
 Finally, we conclude with a brief summary of the results.

The system under study is depicted in Fig. 1(a): Two 1D waveguides, which we
labeled a and b, respectively, chiral coupled to a cascaded three-level
emitter. The cascade three-level system can be a real atom or a manual
atomlike object \cite{51,52,53}. In the circuit quantum electrodynamics system, the
three-level atom coupled to the multiple-modes can be realized by the
Josephson junctions connecting to transmission lines \cite{6,54}. The atomic
states with frequencies $\omega _{1}$, $\omega _{2}$, and $\omega _{3}$ are
denoted by $\left\vert 1\right\rangle $, $\left\vert 2\right\rangle $, and $%
\left\vert 3\right\rangle $. We choose the ground-state energy $\omega _{1}=0
$. In the rotating-wave approximation, the system Hamiltonian has the form $%
\left( \hbar =1\right) $%
\begin{equation}
H=H_{w}+H_{a}+H_{A}+H_{int}.  \label{1-1}
\end{equation}%
The first term $H_{w}$ describing the free propagation of the photons is
given by%
\begin{equation}
H_{w}=-\mathrm{i}\sum_{p=a,b}\int dx\left[ R_{p}^{\dagger }\left( x\right)
\frac{\partial }{\partial x}R_{p}\left( x\right) -L_{p}^{\dagger }\left(
x\right) \frac{\partial }{\partial x}L_{p}\left( x\right) \right] .
\label{1-2}
\end{equation}%
Here we assume that the group velocities of each waveguide are equal and are
set to be $1$ hereafter. And $R_{p}^{\dagger }\left( x\right) \left[
L_{p}^{\dagger }\left( x\right) \right] $ is a bosonic operator creating a
right-going (left-going) photon in waveguide $p=a,b$ at $x$. Representing
the extra cavity of resonance frequency $\omega _{a}$ by the annihilation
operator $a$, its Hamiltonian is
\begin{equation}
H_{a}=\omega _{a}a^{\dag }a.  \label{1-3}
\end{equation}

The third term $H_{A}$ gave the three-level system's Hamiltonian

\begin{equation}
H_{A}=\omega_{2}\sigma_{22}+\omega_{3}\sigma_{33}.  \label{1-4}
\end{equation}
where $\sigma_{mn}=\left\vert m\right\rangle \left\langle n\right\vert
\left( m,n=1,2,3\right) $ represents the atomic energy-level population
operators.

The transition from level $\left\vert 2\right\rangle $ to $\left\vert
3\right\rangle $ is driven by the extra cavity with driving strength $%
\lambda $ and $\omega _{32}=\omega _{3}-\omega _{2}$ is far from the cutoff
frequency of the waveguide. It clearly shows that the waveguides and the
transition $\left\vert 2\right\rangle \leftrightarrow\left\vert
3\right\rangle $ (the extra cavity) are decoupled from each other when $%
\Delta=\omega _{32}-\omega _{k}$ is very large. So the transition $%
\left\vert 2\right\rangle \leftrightarrow\left\vert 3\right\rangle $ doesn't
exchange energy with the waveguides. Moreover, the frequency difference $%
\omega _{32}$ satisfies the conditions $\left\vert \omega _{32}-\omega
_{a}\right\vert \ll\left\vert \omega _{32}+\omega _{a}\right\vert $.
However, the frequency difference between the levels $\left\vert
2\right\rangle $ and $\left\vert 1\right\rangle $ is assumed below the
cutoff frequency of the waveguide, hence, the guided photon is coupled to
the transition $\left\vert 1\right\rangle \leftrightarrow\left\vert
2\right\rangle $ with coupling strength $\gamma_{pg}\left(
p\in\{a,b\},g\in\{r,l\}\right) $. So, the interaction Hamiltonian $H_{int}$
has the form

\begin{align}
H_{int}& =\sum_{p=a,b}\int dx\delta \left( x\right) \left[ \sqrt{\gamma _{pr}%
}R_{p}^{\dagger }\left( x\right) +\sqrt{\gamma _{pl}}L_{p}^{\dagger }\left(
x\right) \right] \sigma _{12}  \notag  \label{1-5} \\
& +\lambda a^{\dag }\sigma _{23}+H.c.
\end{align}
The coupling rates to right- and left-going photons in waveguide $p$ $\left(
p=a,b\right) $ are denoted by $\gamma_{pr}$ and $\gamma_{pl}$, respectively
(Fig. 1(a)). The couplings are chiral as long as $\gamma_{pr}\neq\gamma_{pl}$%
. It is widely known that if the time-reversal symmetry of the waveguide is
broken by a magnetic \cite{55,56,57} or synthetic magnetic field \cite{58,59,60}, then chiral couplings are achieved. When the time-reversal breaking is strong
enough, the ideal chiral waveguide $\left( \gamma_{pg}=\gamma_{p},\gamma_{p%
\bar{g}}=0,\bar{g}\neq g;g,\bar{g}\in\left\{ r,l\right\} \right) $ has been
demonstrated experimentally \cite{55,56,57,58,59,61,62}. The system we are studying belongs
to the chiral QED systems, which have recently attracted widespread
attention \cite{63,64,65,66,67}.

We assumed that a single photon in the waveguide $p$ is incident from the
left side with momenta $k$, the atom is in its ground state $\left\vert
1\right\rangle $, and the extra cavity contains $n$ $\left( \lambda \sqrt{n}%
\ll \left\{ \omega _{32},\omega _{a}\right\} \right) $ photons. Obviously,
when $n=0$, our model is equal to 1D waveguides chiral coupled to a
two-level system \cite{68}. However, when $n\neq 0$, the transition $\left\vert
2\right\rangle \leftrightarrow \left\vert 3\right\rangle $ participates in
the dynamic process, revealing different results from the $n=0$ case.

As a first step, we introduce the parity operators
\begin{subequations}
\label{1-6}
\begin{eqnarray}
E^{\dagger }\left( x\right) &=&\sqrt{\frac{\gamma _{ar}}{\gamma }}%
R_{a}^{\dagger }\left( x\right) +\sqrt{\frac{\gamma _{al}}{\gamma }}%
L_{a}^{\dagger }\left( -x\right) \\
&&+\sqrt{\frac{\gamma _{br}}{\gamma }}R_{b}^{\dagger }\left( x\right) +\sqrt{%
\frac{\gamma _{bl}}{\gamma }}L_{b}^{\dagger }\left( -x\right) ,  \notag
\\
O^{\dagger }\left( x\right) &=&\sqrt{\frac{\gamma _{b}}{\gamma _{a}}}\sqrt{%
\frac{\gamma _{ar}}{\gamma }}R_{a}^{\dagger }\left( x\right) +\sqrt{\frac{%
\gamma _{b}}{\gamma _{a}}}\sqrt{\frac{\gamma _{al}}{\gamma }}L_{a}^{\dagger
}\left( -x\right) \\
&&-\sqrt{\frac{\gamma _{a}}{\gamma _{b}}}\sqrt{\frac{\gamma _{br}}{\gamma }}%
R_{b}^{\dagger }\left( x\right) -\sqrt{\frac{\gamma _{a}}{\gamma _{b}}}\sqrt{%
\frac{\gamma _{bl}}{\gamma }}L_{b}^{\dagger }\left( -x\right) ,  \notag
\end{eqnarray}
\end{subequations}
and
\begin{subequations}
\label{1-7}
\begin{eqnarray}
O_{a}^{\dagger }\left( x\right) &=&\sqrt{\frac{\gamma _{al}}{\gamma _{a}}}%
R_{a}^{\dagger }\left( x\right) -\sqrt{\frac{\gamma _{ar}}{\gamma _{a}}}%
L_{a}^{\dagger }\left( -x\right) ,  \label{1-8} \\
O_{b}^{\dagger }\left( x\right) &=&\sqrt{\frac{\gamma _{bl}}{\gamma _{b}}}%
R_{b}^{\dagger }\left( x\right) -\sqrt{\frac{\gamma _{br}}{\gamma _{b}}}%
L_{b}^{\dagger }\left( -x\right),  \label{1-9}
\end{eqnarray}
\end{subequations}
with $\gamma _{p}=\gamma _{pr}+\gamma _{pl},\gamma =\gamma _{a}+\gamma _{b}$. By employing the above transformation, the original Hamiltonian is transformed into two decoupled Hamiltonians, i.e., $H=H_{e}+H_{o}$, where
\begin{subequations}
\begin{eqnarray}
H_{e} &=&H_{w}^{e}+H_{a}+H_{A}+H_{int}^{e},  \label{1-10} \\
H_{w}^{e} &=&-\mathrm{i}\int dxE^{\dagger }\left( x\right) \frac{\partial }{%
\partial x}E\left( x\right) ,  \label{1-11} \\
H_{int}^{e} &=&\int dx\sqrt{\gamma }\delta \left( x\right) E^{\dagger
}\left( x\right) \sigma _{12}+\lambda a^{\dag }\sigma _{23}+H.c.,
\label{1-12}
\end{eqnarray}
\end{subequations}
and%
\begin{equation}
H_{o}=-\mathrm{i}\int dx\left[ O^{\dagger }\left( x\right) \frac{\partial }{%
\partial x}O\left( x\right) +\sum_{p=a,b}O_{p}^{\dagger }\left( x\right)
\frac{\partial }{\partial x}O_{p}\left( x\right) \right],  \label{1-13}
\end{equation}
with $\left[ H_{e},H_{o}\right] =0$. So, we call $H_{o}$ the interaction-free
independent three-mode Hamiltonian, while $H_{e}$ describes a nontrivial one-mode interacting model with coupling strength $\gamma $.

\begin{figure}[tbph]
\includegraphics[width=10cm]{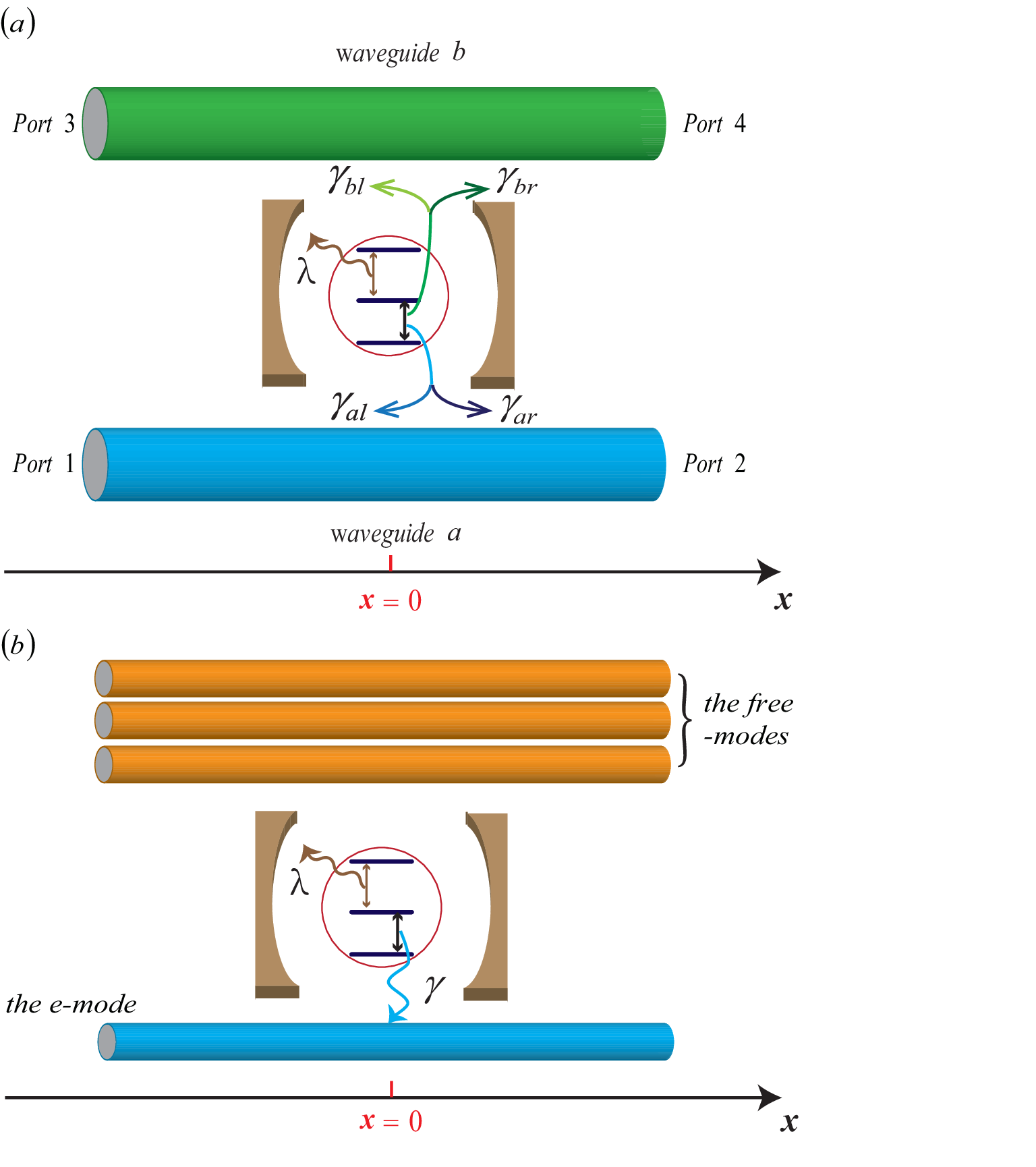}
\caption{(Color online) (a) Schematic of a few-photon router with four ports: A three-level system with cascade configuration interacts with two
independent waveguides labeled $a$ and $b$. The blue (green) line
represents the chiral coupling between the level transitions $\left\vert
1\right\rangle \leftrightarrow \left\vert 2\right\rangle $ and the left and
right-going photons of the bottom waveguide $a$ ($b$), and the coupling
rates are $\protect\gamma _{ar}$ ($\protect\gamma _{br}$) and $\protect%
\gamma _{al}$ ($\protect\gamma _{bl}$), respectively. (b) An equivalent
few-photon router by introducing scatter-free channels and the
controllable channel: The three-level system only interacts with one mode
with a strength determined by $\protect\gamma$.}
\label{fig1}
\end{figure}

The decomposition of Hamiltonian $H$ in Eq. (\ref{1-1}) into two parts
in the previous section leads to the discussion of the problem from one mode
interacting model Hamiltonian $H_{e}$, which greatly
simplifies the calculations. In this section, we will calculate the single-photon scattering process in detail. By the definition of the parity
operators, types Eqs. (\ref{1-6}-\ref{1-7}), $R_{p}^{\dagger }\left( x\right) $
and $L_{p}^{\dagger }\left( x\right) $ can be decomposed into $E^{\dagger
}\left( x\right) $, $O^{\dagger }\left( x\right) $, and $O_{p}^{\dagger
}\left( x\right) .$ Therefore, any one photon state $\left\vert \Psi
^{\left( 1\right) }\right\rangle $\ can be written as
\begin{equation}
\left\vert \Psi ^{\left( 1\right) }\right\rangle =\cos \alpha \left\vert
\Psi \right\rangle _{e}+\sin \alpha \left\vert \Psi \right\rangle _{o}.
\label{1-14}
\end{equation}

One could readily write that the full interacting one-photon eigenstate $%
\left\vert \Psi \right\rangle _{e}$ for $H_{e}$ takes the form%
\begin{eqnarray}
\left\vert \Psi \right\rangle _{e} &=&\int dxf_{k}\left( x\right)
c_{e}^{\dagger }\left( x\right) \left\vert \oslash \right\rangle \left\vert
1,n\right\rangle  \label{1-15} \\
&&+\beta \left\vert \oslash \right\rangle \left\vert 2,n\right\rangle +\zeta
\left\vert \oslash \right\rangle \left\vert 3,n-1\right\rangle ,  \notag
\end{eqnarray}%
where $\left\vert \oslash \right\rangle $ is the vacuum state with no photon
in the two waveguides, and the state $\left\vert m,n\right\rangle \left(
m = 1, 2, 3\right) $ denotes that the atom is in the state $\left\vert
m\right\rangle $ and the extra cavity contains $n$ photons. In addition, $f\left(
x\right) ,$ $\beta $, and $\zeta $ correspond to amplitudes. From the
time-independent Schrödinger equation $H_{e}\left\vert \Psi
_{e}\right\rangle =E_{k}\left\vert \Psi _{e}\right\rangle $, by equating the
coefficients of $c_{e}^{\dagger }\left( x\right) \left\vert \oslash
\right\rangle \left\vert 1,n\right\rangle ,$ $\left\vert \oslash
\right\rangle \left\vert 2,n\right\rangle $, and $\left\vert \oslash
\right\rangle \left\vert 3,n-1\right\rangle $, respectively, we obtain the
equations for the amplitudes
\begin{subequations}
\label{1-16}
\begin{eqnarray}
\left( -\mathrm{i}\frac{\partial }{\partial x}-\delta _{k}^{n}\right)
f_{k}\left( x\right) +\sqrt{\gamma }\beta \delta \left( x\right) &=&0, \\
-\Delta _{k}^{n}\beta +\sqrt{\gamma }f_{k}\left( 0\right) +\sqrt{n}\lambda
\zeta &=&0, \\
-\left( \Delta _{k}^{n}+\Delta _{a}\right) \zeta +\sqrt{n}\lambda \beta &=&0.
\end{eqnarray}%
Where $\delta _{k}^{n}=E_{k}-n\omega _{a}$ represents the eigenfrequency of
the incident single photon, i.e., $vk$, and the detuning $\Delta
_{k}^{n}=\delta _{k}^{n}-\omega _{2},$ $\Delta _{a}=\omega _{a}-\omega _{32}$
describes the energy difference between the eigenfrequency of the incident
single photon $vk$ and $\left\vert 2\right\rangle \leftrightarrow \left\vert
1\right\rangle ,$ the extra cavity frequency $\omega _{a}$ and $\left\vert
3\right\rangle \leftrightarrow \left\vert 2\right\rangle $, respectively.
Eq. (\ref{1-16}c) can be rewritten as
\end{subequations}
\begin{equation}
\zeta =\frac{\sqrt{n}\lambda }{\Delta _{k}^{n}+\Delta _{a}}\beta ,
\label{1-17}
\end{equation}%
substituting Eq. (\ref{1-17}) into Eq. (\ref{1-16}b) yields%
\begin{equation}
\beta =\sqrt{\gamma }\delta \left( x\right) \frac{\left( \Delta
_{k}^{n}+\Delta _{a}\right) }{\Delta _{k}^{n}\left( \Delta _{k}^{n}+\Delta
_{a}\right) -n\lambda ^{2}}f_{k}\left( x\right) ,  \label{1-18}
\end{equation}%
again, applying the constraint Eq. (\ref{1-18}) to Eq. (\ref{1-16}a) yields%
\begin{equation}
\left[ \left( -\mathrm{i}\frac{\partial }{\partial x}-\delta _{k}^{n}\right)
+V\delta \left( x\right) \right] f_{k}\left( x\right) =0,  \label{1-19}
\end{equation}%
where we have introduced the energy-dependent $\delta $-like potentials $%
V\equiv \gamma \frac{\left( \Delta _{k}^{n}+\Delta _{a}\right) }{\Delta
_{k}^{n}\left( \Delta _{k}^{n}+\Delta _{a}\right) -n\lambda ^{2}}$. It is
clear that when $n=0$, the potential is the result in Ref. \cite{46}, at
the incident energy resonance, i.e., $\Delta _{k}^{n}=0$, $V$ goes to
infinity. Notice that if the a-mode contains photons, that is, $n\neq 0$,
but the a-mode photon resonantly drives the atomic transition, i.e., $\Delta
_{a}=0$, the potential equals zero. More broadly speaking, if $\Delta
_{k}^{n}+\Delta _{a}=0,$ then $V\rightarrow 0;$ if $\Delta
_{k}^{n}\left( \Delta _{k}^{n}+\Delta _{a}\right) -n\lambda ^{2}=0$, then $V\rightarrow \infty $. Therefore, we can
regulate the effective potential $V$ from $0$ to $\infty $ by the parameters
of the a-mode field. Obviously, the plane-wave solution $f_{k}\left(
x\right) $ is sufficient to satisfy Eq. (\ref{1-21})
\begin{eqnarray}
f_{k}\left( x\right) &=&P_{k}\left( x\right) \left[ \Theta \left( -x\right)
+t_{k,e}\Theta \left( x\right) \right] ,  \label{1-20} \\
t_{k,e} &=&\frac{\Delta _{k}^{n}-\frac{n\lambda ^{2}}{\Delta _{k}^{n}+\Delta
_{a}}-\mathrm{i}\frac{\gamma }{2}}{\Delta _{k}^{n}-\frac{n\lambda ^{2}}{%
\Delta _{k}^{n}+\Delta _{a}}+\mathrm{i}\frac{\gamma }{2}}.  \label{1-21}
\end{eqnarray}%
In all the following calculations, we set $f_{k}\left( 0\right) =\frac{1}{2}%
\left[ f_{k}\left( 0^{+}\right) +f_{k}\left( 0^{-}\right) \right] $ \cite%
{69,70}, and $\Theta \left( x\right) $ is the step function. And we
assume that in the incident interval, the incoming wave functions consist
entirely of plane waves \cite{32,69,70}, that is, for $x<0$, $%
f_{k}\left( x\right) =P_{k}\left( x\right) =\frac{1}{\sqrt{2\pi }}e^{\mathrm{%
i}kx}.$

Note that $t_{k,e}$ is the transmission coefficient for the interacting
model because the even mode is extremely chiral, $\left\vert
t_{k,e}\right\vert ^{2}=1.$ In the actual operation, for a normalized state
\begin{equation}
\left\vert \Psi _{in}\right\rangle _{k}=\left\vert \Psi _{in}\right\rangle
_{G_{p}}=\int dx\frac{1}{\sqrt{2\pi }}e^{\mathrm{i}kx}G_{p}^{\dagger }\left(
x\right) \left\vert \oslash \right\rangle \left\vert 1,n\right\rangle ,
\label{1-22}
\end{equation}%
which considers a monochromatic right-going $\left( G=R\right) $ or left-going $%
\left( G=L\right) $ photon incident from the waveguide $p$, the out-state is

\begin{eqnarray}
\left\vert \Psi _{out}\right\rangle _{k} &=&t_{pg,k}^{p}\left\vert \Psi
_{out}\right\rangle _{G_{p}}+r_{pg,k}^{p}\left\vert \Psi _{out}\right\rangle
_{\bar{G}_{p}} \\
&&+t_{pg,k}^{\bar{p}}\left\vert \Psi _{out}\right\rangle _{G_{\bar{p}%
}}+r_{pg,k}^{\bar{p}}\left\vert \Psi _{out}\right\rangle _{\bar{G}_{\bar{p}%
}}.  \notag  \label{1-23}
\end{eqnarray}%
where the four-mode transmission amplitudes $t_{pg,k}^{p},t_{pg,k}^{\bar{p}},$
and reflection amplitudes $r_{pg,k}^{p},r_{pg,k}^{\bar{p}}$ are
\begin{subequations}
\label{1-24}
\begin{eqnarray}
t_{pg,k}^{p} &=&\frac{\gamma _{pg}}{\gamma }U_{k}+1, \\
r_{pg,k}^{p} &=&\frac{\sqrt{\gamma _{pg}\gamma _{p\bar{g}}}}{\gamma }U_{k},
\\
t_{pg,k}^{\bar{p}} &=&\frac{\sqrt{\gamma _{pg}\gamma _{\bar{p}g}}}{\gamma }%
U_{k}, \\
r_{pg,k}^{\bar{p}} &=&\frac{\sqrt{\gamma _{pg}\gamma _{\bar{p}\bar{g}}}}{%
\gamma }U_{k}.
\end{eqnarray}%

To simplify, we introduce the scattering factors $U_{k}$, expressed
as
\end{subequations}
\begin{equation}
U_{k}\equiv t_{k,e}-1=\frac{-\mathrm{i}\gamma \left( \Delta _{k}^{n}+\Delta
_{a}\right) }{\left( \Delta _{k}^{n}\left( \Delta _{k}^{n}+\Delta
_{a}\right) -n\lambda ^{2}\right) +\mathrm{i}\frac{\gamma }{2}\left( \Delta
_{k}^{n}+\Delta _{a}\right) }.  \label{1-25}
\end{equation}

\begin{figure*}[tbph]
\includegraphics[width=16cm]{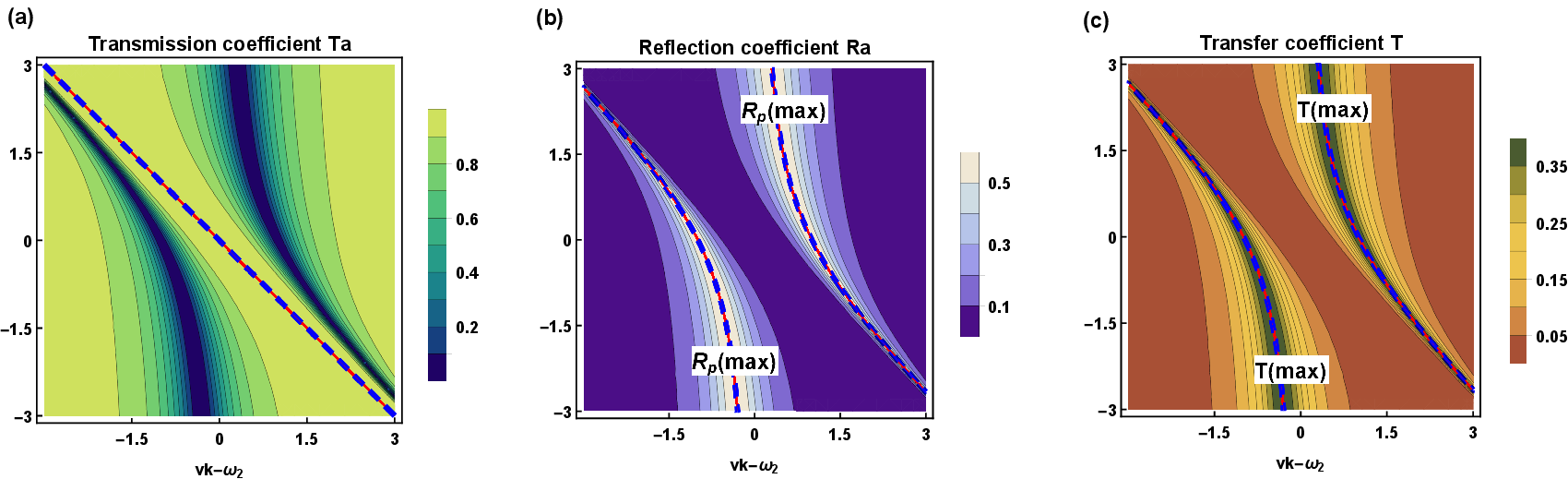}
\caption{(Color online) Single photon routing probability: (a) Transmission, (b)
Reflection, and (c) Transfer coefficients versus the detuning $\Delta
_{k}^{n} $ and $\Delta _{a}$. We set $\protect\gamma _{pg}/\protect%
\gamma =1/2,\protect\gamma _{p\bar{g}}/\protect\gamma =0.3,\protect\gamma _{%
\bar{p}g}/\protect\gamma =\protect\gamma _{\bar{p}\bar{g}}/\protect\gamma %
=0.1,\protect\lambda /\protect\gamma =1$, and $n=1$. All the parameters but n are
in units of $\protect\gamma $.}
\label{fig2}
\end{figure*}

\begin{figure}[tbph]
\includegraphics[width=8cm]{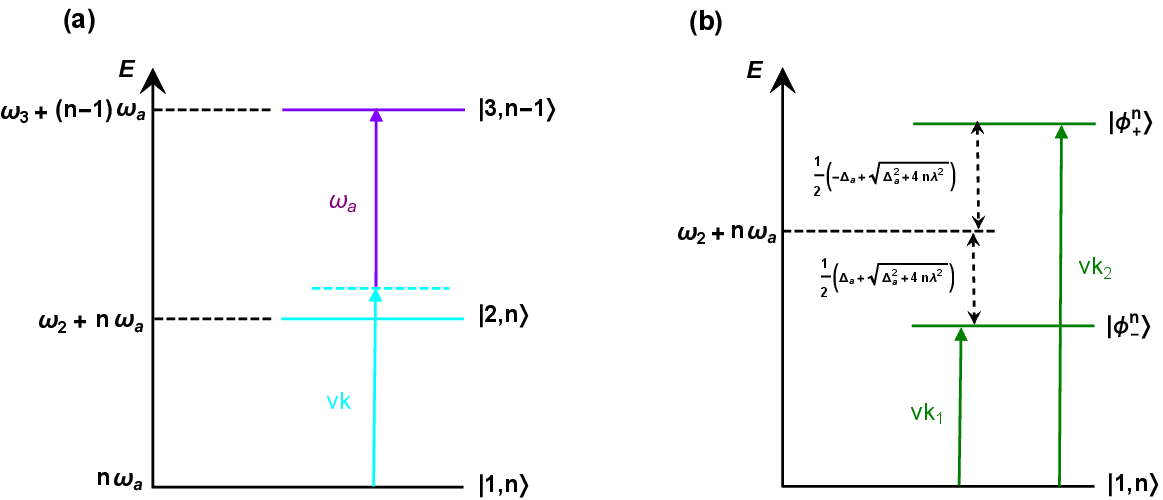}
\caption{(Color online) The energy-level diagram shows a single incident
photon-induced tunneling (PIT) in (a), and in the dressed state representation, it can
be considered an effective V-type atom with the energy-level diagram
shown in (b). }
\label{fig3}
\end{figure}

The corresponding four scattering coefficients are defined as $%
T_{p}\equiv \left\vert t_{pg,k}^{p}\right\vert ^{2}$, $R_{p}\equiv
\left\vert r_{pg,k}^{p}\right\vert ^{2}$, $T_{\bar{p}}\equiv
\left\vert t_{pg,k}^{\bar{p}}\right\vert ^{2}$, $R_{\bar{p}}\equiv
\left\vert r_{pg,k}^{\bar{p}}\right\vert ^{2}$. $T_{p}$\ and $%
R_{p}$\ are the transmission and reflection coefficients of waveguide %
$p$; $T_{\bar{p}}$\ and $R_{\bar{p}}$\ are the
transmission and reflection coefficients of waveguide $\bar{p}$; $%
\bar{p}\neq p$. If the photon is transferred from waveguide $p$%
\ to waveguide $\bar{p}$, the transfer coefficient is denoted by $T$%
, which is $T\equiv T_{\bar{p}}+R_{\bar{p}}$. The single-photon
scattering probabilities $T_{p}$, $R_{p}$\ , and $T$\ with
respect to detuning $\Delta _{a}$\ and $\Delta _{k}^{n}$\ are shown
in Fig. 2.  The red-blue asymptotic line in Fig. \ref{fig2}(a) implies that the incident photon completely transmits ($T_{p}= 1$) for $\Delta _{k}^{n}+\Delta _{a}=0$. This is because the
incident photon induces the photon transition in the extra cavity, causing the atom to directly transition to the highest state $\left\vert3\right\rangle$, i.e., $\omega _{3}=\omega _{a}+vk,$ resulting in photon-induced tunneling (PIT), as shown in Fig. \ref{fig3}(a).
Besides, the red-blue asymptotic line in Fig. \ref{fig2}(b)-(c) represents the incident photon that can't transmit ($T_{p}=0$), and the reflect coefficient ($R_{p}$) and transfer coefficient ($T$) reach their maximum for $\Delta _{k}^{n}\left( \Delta _{k}^{n}+\Delta
_{a}\right) -n\lambda ^{2}=0$. It can be seen that the routing probability attains its maximum $R_{p}=60$\%$ $ and $T=40$\%$ $ at two red-blue asymptotic lines. Generally, the maximum routing
probability (i.e., the routing peak) corresponds with the resonant input photons
(i.e., $\Delta _{k}^{n}=0$) \cite{46}. The splitting and shifting of the
routing peaks in Fig. \ref{fig2}(b)-(c) show the effects of the Rabi splitting, which
is caused by the coupling between the extra cavity and the atom. It may be better
understood by substitution of the dressed state, that is, under $n$
excitation subspaces, the coupling model of the extra cavity mode $a$ and
atom $\left\vert 3\right\rangle \leftrightarrow \left\vert 2\right\rangle $
is equivalent to $H_{p}^{n}=E_{+}^{n}\left\vert \phi _{+}^{n}\right\rangle
\left\langle \phi _{+}^{n}\right\vert +E_{-}^{n}\left\vert \phi
_{-}^{n}\right\rangle \left\langle \phi _{-}^{n}\right\vert .$ Thus, our
emitter can be considered a V-type system in the dressed state
representation, as shown in Fig. \ref{fig3}(b). The effective transition $%
\left\vert 1,n\right\rangle \leftrightarrow \left\vert \phi _{\pm
}^{n}\right\rangle $ can be derived.

Fig. 3(a) shows the structure of the quantum emitter. The state $%
\left\vert 3,n-1\right\rangle $ does not appear when the initial state of
the extra cavity is in a vacuum state and the atom is in the ground state. At this
time, it's impossible for the atom to transition from the state $\left\vert
2\right\rangle $ to the state $\left\vert 3\right\rangle $, and then the
frequency corresponding to this particular tunnel route does not appear. To
put it another way, under the dressed state appearance, as
shown in Fig. 3(b), only the dressed state $\left\vert \phi
_{+}^{n}\right\rangle $ plays a role at this point, and the emitter is
equivalent to a two-level atom. Note that the tunneling routing points in
Fig. 3(a) must exist once the extra cavity contains photons, while Fig. 3(b)
needs to be discussed in detail. The Rabi splitting $\Omega _{n}\left( \Delta
_{a}\right) =\sqrt{\Delta _{a}^{2}+4n\lambda ^{2}}$ indicates that the single-photon transport properties are influenced by those parameters of the
Jaynes-Cummings-like model, such as the detuning $\Delta _{a}$,
atom-cavity coupling strengths $\lambda $, and the number of photons $n$ in
the extra cavity.
\begin{figure*}[tbph]
\includegraphics[width=18cm]{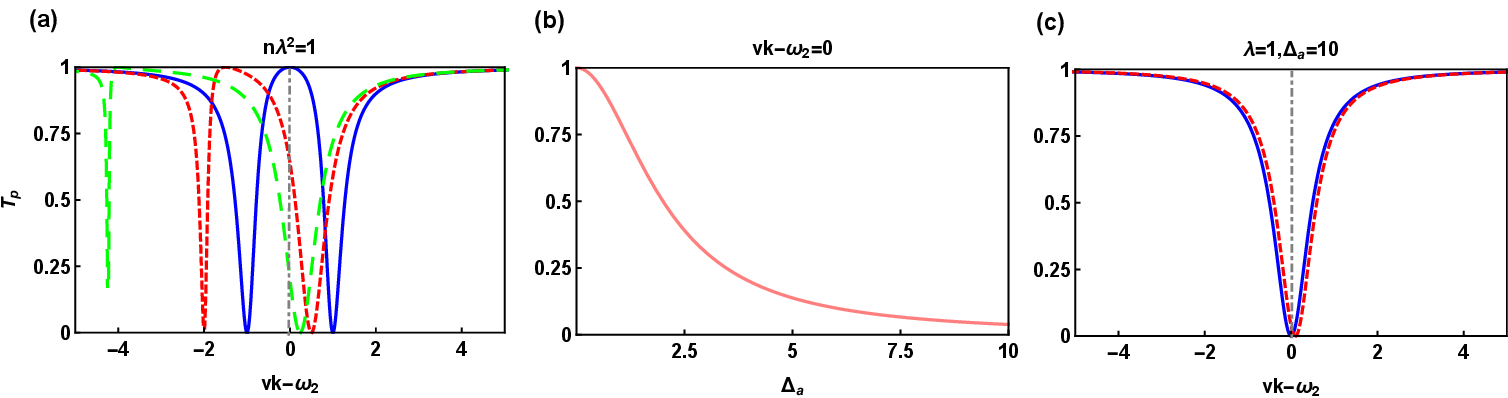}
\caption{(Color online) Transmission of a single photon against the detuning
$vk-\protect\omega _{2}$ in (a), (c), and against the detuning $\Delta _{a}$
in (b). The blue solid, red dashed, and green long dashed lines in panels
(a) denote that the detuning $\Delta _{a}$ is taken as 0, 1.5, and 4,
respectively. The black dotted dashed line in the middle of (a) represents
the intersection of the three conditions when the incident resonant
frequency. (b) is the slice diagram at this time. We consider $n=1$ in
(a), (b), and (c), where $n=0 $ (blue solid line) and $n=1 $ (red dashed line). Other
parameters are set as follows: $\protect\gamma _{pg}/\protect\gamma =1/2, %
\protect\gamma _{p\bar{g}}/\protect\gamma =0.3, \protect\gamma _{\bar{p}g}/%
\protect\gamma =\protect\gamma _{\bar{p}\bar{g}}/\protect\gamma =0.1, \protect%
\lambda /\protect\gamma =1$. All the parameters but $n $ are in units of $%
\protect\gamma $. }
\label{fig4}
\end{figure*}

In Fig. \ref{fig4}, we consider that $\lambda $ and $\gamma $ have the same
order of magnitude, such as $\lambda =\gamma $. In Fig. 4(a), we plot the transmission spectra of a single photon at $\Delta
_{a}=0$, $1.5$, and $4$, respectively. The figure shows that each
value $\Delta _{a}$ has two different valleys and that the value determines the spectrum's symmetric or asymmetric properties. When $\Delta _{a}=0$, $T_{p}$ can show symmetry as a function of $vk-\omega _{2}$.
 With the increase of $\Delta _{a}$, the symmetry of
the spectrum is broken. The left valley shifts to the left, and the line
width becomes narrower and sharper. Also, the right valley shifts to
the left, infinitely close to zero, but does not cross zero on the left axis.
Although the line width of the right valley increases with increasing $%
\Delta _{a}$, it does not exceed the limit $\gamma /2$, which is the
decay width of the atom. In addition, Fig. 4(b) shows that the larger the $%
\Delta _{a}$, the lower the transmittance at the resonance. When $\Delta
_{a}$ is large enough, that the number of photons in the extra cavity has almost
no effect on the single-photon transmittance because the extra cavity is
almost decoupled from the atom at this point, as shown in Fig. 4(c). All of this can be explained by the dressed states in
Fig. \ref{fig3}(b), it is well known that in the large-detuning case,
i.e., $\Delta _{a}\gg \sqrt{n}\lambda.$ The dressed states $\left\vert \phi
_{\pm }^{n}\right\rangle \ $can be approximately written as:$\ \left\vert
\phi _{+}^{n}\right\rangle \approx \left\vert 2,n\right\rangle -\frac{\sqrt{n%
}\lambda }{\Delta _{a}}\left\vert 3,n-1\right\rangle ;\left\vert \phi
_{-}^{n}\right\rangle \approx \left\vert 3,n-1\right\rangle +\frac{\sqrt{n}%
\lambda }{\Delta _{a}}\left\vert 2,n\right\rangle $ \cite{71}. The corresponding effective coupling strengths are $\gamma _{+}^{n}\approx \frac{\Delta
_{a}^{2}}{\Delta _{a}^{2}+n\lambda ^{2}}\gamma ,\gamma _{-}^{n}\approx \frac{%
\sqrt{n}\lambda }{\Delta _{a}}\gamma _{+}^{n}\approx \sqrt{n}\lambda \frac{%
\Delta _{a}}{\Delta _{a}^{2}+n\lambda ^{2}}\gamma $, respectively. Therefore, as the $\Delta _{a}$ increases, the line width of the left
valley becomes sharper and sharper, and the line width of the right valley
becomes closer and closer to $\gamma /2$. In the $\frac{\Delta _{a}}{%
\sqrt{n}\lambda }$ $\rightarrow \infty $ limit, it is difficult for the
system to transition from the state $\left\vert 2,n\right\rangle $ to the state $\left\vert 3,n-1\right\rangle $. At this point, there is no
significant difference in whether the extra cavity contains photons or not,
and the system is equivalent to a two-level emitter. Of course, all of these values always stayed within
the rules of $\left\vert \Delta _{a}\right\vert \ll \omega _{a}+\omega
_{32}, $ and $\sqrt{n}\lambda \ll \min \left\{ \omega _{a},\omega
_{32}\right\} $. Then one obtains the similar Jaynes-Cummings Hamiltonian $%
\omega _{a}a^{\dag }a+\omega _{2}\sigma _{22}+\omega _{3}\sigma
_{33}+\lambda a^{\dag }\sigma _{23}+H.c..$

Fig. 4 shows the property of the single-photon transmission spectrum for $%
\Delta _{a}>\sqrt{n}\lambda $, while Fig. 5 shows the change that occurs
in the transmission spectrum for $\Delta _{a}<\sqrt{n}\lambda $. As shown
in Fig. 5(a), by increasing the number of photons $n$ in the extra cavity (expanding Rabi splitting), the window of the transmission spectrum $T_{p}$
is widened, and the spectral line shape tends to be symmetrical. In Fig.
5(b), we compare the different transmissions that vary with $n$ when $%
\Delta _{a}=1$, $5$, and $10$. The figure shows that
large transmission can be achieved with a tiny $\Delta _{a}$ and a larger
number of photons, $n$. This means that under these values, the effective
coupling strength $\sqrt{n}\lambda $ of the transition $\left\vert
2,n\right\rangle \leftrightarrow \left\vert 3,n-1\right\rangle $ is large
enough, and the detuning $\Delta _{a}$ between the bare states $\left\vert
2,n\right\rangle $ and $\left\vert 3,n-1\right\rangle $ is small enough, so the
transition probability between $\left\vert 2,n\right\rangle \leftrightarrow
\left\vert 3,n-1\right\rangle $ is much greater than the transition
probability between $\left\vert 2,n\right\rangle \leftrightarrow \left\vert
1,n\right\rangle $. As for the case where the splitting width is less than
the decay rate of the emitter, i.e., $\Omega _{n}\left( \Delta _{a}\right)
<\gamma /2,$ we will not elaborate here because the two decorated states
are very close to each other, putting higher requirements on
experimental observation.

\begin{figure*}[tbph]
\includegraphics[width=16cm]{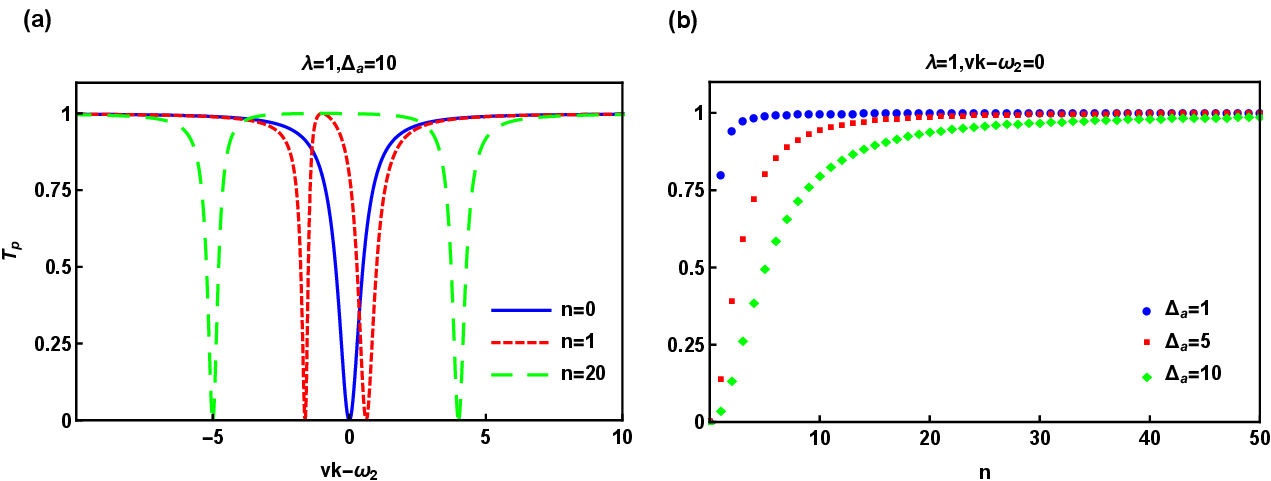}
\caption{(Color online) Transmission probability $T_{p}$ as a function of $%
vk-\protect\omega _{2}$ for three different initial extra cavity
photon states in (a). State-1: vacuum state (blue solid line); State-2: single photon state (red dashed line); State-3: 20-photon state (green
long dashed line). We consider $\Delta _{a}=\protect\lambda =\protect\gamma $
in panel (a). $T_{p}$ as a function of $n$ for $\Delta _{a}=1$ (blue point
symbol), $\Delta _{a}=5$ (red square symbol), $\Delta _{a}=10$ (green
rhombus symbol) in panel (b), and the energy of the incident photon resonates
with the emitter, i.e., $vk-\protect\omega _{2}=0$. The other parameters
remain the same as in Fig. \protect\ref{fig4}. }
\label{fig5}
\end{figure*}

In this paper, we have studied the single-photon routing in two 1D linear waveguide systems coupled with a three-level emitter with a
cascade configuration. An extra cavity photon can drive the atomic transition from $\left\vert 2\right\rangle $ to $\left\vert 3\right\rangle $. So, if the emitter is in the ground state $\left\vert 1\right\rangle $ initially, whether the cavity
contains photons determines the effective configuration of the atom. Here, we have shown that complete transmission occurs if the extra cavity applied to the 3LS initially contains photons when $\Delta
_{k}^{n}+\Delta _{a}=0$. Besides, the routing peaks splitting and shifting in Fig. \ref{fig4}(a) show the influence of the
Rabi splitting, which is caused by the coupling between the extra cavity and the atomic transition $\left\vert
2\right\rangle \leftrightarrow \left\vert 3\right\rangle $. In other words, we obtain an ATS line shape, and the transparency window can be controlled by the number of photons in the extra cavity. In addition, the location and width of the transmission valleys are tunable by adjusting the extra cavity.
Consequently, the single-photon transport property depends on whether the extra cavity contains a photon or not. The single-photon transport can also be controlled by the number of photons in the extra cavity. Therefore, the proposal presented in this paper can be a competitive candidate for quantum routes. We hope that our proposal can provide a feasible approach to
constructing various photonic networks and be used to do some further research, such as directional routing of a single photon with different frequencies and so on.

\begin{acknowledgments}
This work was supported by NSFC Grants No.11975095, No.12075082, No.11935006,
and the Science and Technology Innovation Program of Hunan Province (Grant No. 2020RC4047)
\end{acknowledgments}

\section*{Data Availability}
The data that support the findings of this study are available from the corresponding author upon reasonable request.

\nocite{*}
\bibliography{mybibtex}% Produces the bibliography via BibTeX.

\end{document}